\documentclass[letter]{aa}
\usepackage{natbib,twoopt}
\usepackage[breaklinks=true]{hyperref} 
\bibpunct{(}{)}{;}{a}{}{,}             
\makeatletter
  \newcommandtwoopt{\citeads}[3][][]{\href{http://adsabs.harvard.edu/abs/#3}%
    {\def\hyper@linkstart##1##2{}%
     \let\hyper@linkend\@empty\citealp[#1][#2]{#3}}}
  \newcommandtwoopt{\citepads}[3][][]{\href{http://adsabs.harvard.edu/abs/#3}%
    {\def\hyper@linkstart##1##2{}%
     \let\hyper@linkend\@empty\citep[#1][#2]{#3}}}
  \newcommandtwoopt{\citetads}[3][][]{\href{http://adsabs.harvard.edu/abs/#3}%
    {\def\hyper@linkstart##1##2{}%
     \let\hyper@linkend\@empty\citet[#1][#2]{#3}}}
  \newcommandtwoopt{\citeyearads}[3][][]%
    {\href{http://adsabs.harvard.edu/abs/#3}
    {\def\hyper@linkstart##1##2{}%
     \let\hyper@linkend\@empty\citeyear[#1][#2]{#3}}}
\makeatother
\usepackage{graphicx}
\usepackage{txfonts}
\usepackage{natbib}
\newcommand{\doce}{\mbox{$^{12}$CO}}

\newcommand{\trece}{\mbox{$^{13}$CO}}

\newcommand{\htrececn}{\mbox{H$^{13}$CN}}
\newcommand{\jkk}{\mbox{$J_{K_{\rm a},K_{\rm c}}$}}
\newcommand{\jkkul}[6]{\mbox{$#1_{#2,#3}$--$#4_{#5,#6}$}}
\newcommand{\jos}{\mbox{$J$=8$-$7}}

\newcommand{\jct}{\mbox{$J$=4$-$3}}
\newcommand{\jtd}{\mbox{$J$=3$-$2}}

\newcommand{\kms}{\mbox{km\,s$^{-1}$}}

\newcommand{\mJybeam}{\mbox{mJy\,beam$^{-1}$}}
\newcommand{\ms}{\mbox{$M_{\mbox{\sun}}$}}

\newcommand{\lsim}{\raisebox{-.4ex}{$\stackrel{\sf <}{\scriptstyle\sf \sim}$}}
\newcommand{\gsim}{\raisebox{-.4ex}{$\stackrel{\sf >}{\scriptstyle\sf \sim}$}}

\newcommand{\farcss}{\mbox{\rlap{.}$''$}}

\begin{document}

   \title{The very compact dust disk in the Red Rectangle}

   \author{V. Bujarrabal
          \inst{1}
            \and J. Alcolea\inst{2}
          \and
          A. Castro-Carrizo\inst{3}
          \and
J. Kluska\inst{4}
          \and
C. Sánchez Contreras\inst{5}
          \and
H. Van Winckel\inst{4}  }

   \institute{             Observatorio Astron\'omico Nacional (OAN-IGN),
              Apartado 112, E-28803 Alcal\'a de Henares, Spain\\
              \email{v.bujarrabal@oan.es}
\and
             Observatorio Astron\'omico Nacional (OAN-IGN),
             C/ Alfonso XII, 3, E-28014 Madrid, Spain
\and
 Institut de Radioastronomie Millim\'etrique, 300 rue de la Piscine,
 38406, Saint Martin d'H\`eres, France
\and
Instituut voor Sterrenkunde, KU Leuven, Building D, Celestijnenlaan 200B, 3001
Leuven, Belgium
          \and
Centro de Astrobiología (CAB), CSIC-INTA, ESAC Campus, E-28692 Villanueva de la Cañada, Madrid, Spain}

   \date{submitted 9 August 2023; accepted 6 September 2023}

  \abstract
   {}
   {We aim to study the dust distribution in the central regions of the Keplerian disk of the Red Rectangle, the prototype of binary post-AGB stars with rotating circumbinary disks, and to compare it with the distribution of relevant molecular gas tracers.}
{We present new high-resolution (20 milliarcseconds, mas) ALMA observations of continuum and line emissions at 0.9\,mm. The maps have been analyzed by means of a simple model of dust and free-free emission that is able to reproduce the continuum data. }
{i) We find that most of the dust emission in the Red Rectangle is concentrated in the inner disk regions, with a typical size of 250 AU in diameter and 50 AU in width. ii) The settlement of dust grains onto inner equatorial regions is remarkable when compared with the relatively widespread gas distribution. iii) This region is basically coincident with the warm PDR (photo-dominated region) where CI, CII, and certain molecules such as HCN are presumably formed, as well as probably PAHs (polycyclic aromatic hydrocarbons, whose emission is very strong in this source). iv) We confirm the large size of the grains, with a typical radius $\sim$ 150 $\mu$m. The opacity of dust at 0.9 mm is deduced to be relatively large, $\sim$0.5.
v) We also confirm the existence of a very compact HII region in the center, for which we measure an extent of 10 -- 15 mas ($\sim$ 10 AU) and a total flux of 7 -- 8 mJy at 0.9\,mm.}
  {}


   \maketitle
%

\section{Introduction}

The Red Rectangle is a prototype of binary post-AGB stars surrounded by Keplerian or quasi-Keplerian circumbinary disks. About 85 objects in our Galaxy have been classified into this ample class \citep[e.g.,][]{win03,kluska22}. Their spectral energy distributions (SEDs) are peculiar, with a near-infrared (NIR) excess indicative of hot dust close to the stellar system. The IR interferometry of such inner dust emission shows tiny ring-like structures \citep{deroo06, hillen16, kluska19}, where the dust reaches $\sim$1000 K, the highest temperature that can be reached before grain sublimation is initiated. This suggests a stable circumstellar structure, as the stars are no longer in a state of copious mass loss \citep{ruyter06, gezer15}. The  high degree of dust processing observed \citep[e.g.,][]{gielen11} provides additional indications of the longevity of these disks. Single-dish observations of \doce\ and \trece\ mm-wave emission in a significant sample of these post-AGB stars have systematically yielded characteristic line profiles, with a prominent single or double peak and moderate-velocity wings \citep{bujetal13a}, which are known to be very reliable indicators of the existence of rotating disks. Rotation has been confirmed in several objects by high-resolution CO-line mapping \citep[e.g.,][]{bujetal15, bujetal16}. In the Red Rectangle, a region in which chemistry is dominated by photodissociation (PDR) was also identified in very inner  disk regions from the emission of characteristic molecular and atomic lines and polycyclic aromatic hydrocarbons (PAHs), \cite{bujetal16}, probably created by the high-energy spectrum of an accretion disk surrounding the secondary \citep{thomas13}.

The Red Rectangle consists of a double stellar system with an orbital period of about 320 days, surrounded by an extended disk (with a diameter $\sim$ 6$\times$10$^{16}$ cm) and a beautiful X-shaped axisymmetric nebula in expansion, seen in radio and optical/NIR images \citep[e.g.,][]{cohen04, bujetal13b, bujetal16}.  The  axis of symmetry and equator are conspicuous; the axis is slightly out of the plane of the sky, by about 5$^\circ$, and its projection in the plane of the sky shows PA $\sim$ 10 -- 15 degrees. \cite{men02} deduced a distance of 710 pc for this source, which is compatible with our previous analysis of CO maps and likewise adopted in this letter. The presence of a dominant population of big grains in the Red Rectangle, over 0.1 mm in size, was proposed by \cite{jura95, jura97}. These authors also detected free-free emission from a very compact HII region at the center of the nebula, whose cm-wave emission shows fluxes $\sim$ 0.5 -- 1 mJy and extends $<$ 0\farcss 1.

The modeling of the overall dust emission and scattering by \cite{men02} indicated the presence of a dense and hot dust disk with two distinct regions. The innermost one extends up to $\sim$ 1.1$\times$10$^{15}$ cm (73 AU) and shows a relatively constant density and a temperature of 300 -- 1000 K. The second one is a less dense plateau reaching about 5$\times$10$^{15}$ cm. Farther out, the density decreases gradually but very steeply. There is also a central cavity with a radius of 14 AU. These general properties are similar to those found from our mapping and modeling, despite the fact
that the previous work lacked empirical information on the mm-wave emission and disk dust distribution.
Excusably, their model didn't include the strong settlement of dust towards the equator shown in our data.

\section{Observations}

\subsection{Data reduction}

We present results from ALMA
  project 2019.1.00177.S. 
These observations were carried out in band 7 in two array configurations, between July and September 2021.
Data were obtained with baselines ranging from 14\,m to 14.9\,km, for the spectral windows 
330.350--330.827\,GHz (at 0.21\,\kms\ resolution), 330.744--332.658\,GHz (0.85\,\kms),
343.563--345.448 (0.85\,\kms), and  345.558--346.032 (0.21\,\kms). Standard observations were made
alternating acquisitions on source and several radio quasars for calibration purposes. In particular, the reference flux calibrator was J0522-3627 (with flux 4.37 Jy at 343.5 GHz, and
spectral index $-$0.27) to derive a flux of 0.504 Jy for the phase calibrator J0607-0834. The data
were processed with the standard calibration pipeline in the CASA software package (version 6.1.1.15).
After calibration, the science target data were exported for further processing into the GILDAS software package\footnote{GILDAS is a software package to analyze single-dish and interferometric millimeter data. More information can be found in https://www.iram.fr/IRAMFR/GILDAS/}.
Line-subtracted continuum emission data were used to self-calibrate phases (with simple 
approximations and standard programs) in all the datasets, which allowed for
the  noise to be reduced close to the expected thermal noise level.

We performed image synthesis for all the lines and continuum data per
spectral window, using the Hogbom method and several types of visibility weighting. In Fig.\,1, we present 
the obtained results for the continuum by averaging all spectral windows. We convolved  
the clean components, derived with a robustness factor of 0.2 (in the GILDAS scheme), with a circular beam of 20 mas (milliarcseconds),
very close to the original synthetic beam of 21$\times$19 mas at PA 65$^\circ$.
The continuum emission consists of a central compact component plus the contribution of an edge-on
disk. The central component, probably due to free-free emission (Sects.\ 1 and 4), extends 10 -- 15 mas and shows a flux of 8$\pm$2 mJy, as derived by only considering long-baseline ($>$ 5 km) visibilities. The J2000 ICRS coordinates of the
brightness peak are R.A. 06$^{\rm h}$19$^{\rm m}$58\fs1988 and Dec. $-$10\degr38\arcmin15\farcs216, which have
been adopted as the center of the nebula and origin of coordinates in all the maps presented here.

Data were also obtained for the emission of \doce\ and \trece\ \jtd, and \htrececn\ \jct, as well as the vibrationally 
excited lines \htrececn\ $\nu_2$=1 \jct\ $l$=1$e$, H$_2$O $\nu_2$=2 \jkk=\jkkul{3}{2}{1}{4}{1}{4},
and SiO $\nu$=1 \jos. Image synthesis was made having subtracted the continuum in the {\em uv}-plane and
adopting natural weighting for better sensitivity, which led to synthetic beams of about
61$\times$46 mas at PA $-$69\degr. Line maps will be discussed in detail in a forthcoming paper and only some data are shown in this letter.

\subsection{Calibration of flux and representative spectral index}

We have analyzed in detail the flux calibration in our maps and 
the amount of flux that could have been filtered out in these interferometric data. Upon checking the adopted fluxes for flux calibrators in both datasets, we do not expect uncertainties larger than 10\%. The total integrated flux in the new 0.9-mm continuum maps is $\sim$ 0.66 Jy. This is almost identical to the flux reported by \cite{bujetal13b}, $\sim$ 0.65 Jy, with a very different array configuration (a synthetic beam $\sim$ 30 times larger; maps performed in 2012). In addition, by analyzing the data in the {\em uv}-plane we have checked that in both datasets we got baselines that are similarly short, and that the amount of flux in those visibilities is also very similar. We want to underline the very high dynamic range of the previous observations. The first contour in them  extends $\sim$ 2$''$ along the equator, at a level of 1.5 \mJybeam\ (0.05 K), about 500 times lower than the central peak: those very extended regions only represent about 10\% of the total flux, which mostly comes from the inner 0\farcs 2, practically unresolved in the 2012 maps. Even if this contour is not detected at all in our data, the continuum flux lost by overresolution would be \lsim\ 10\%.

   \begin{figure}[h]
   \rotatebox{0}{\resizebox{9cm}{!}{
   \includegraphics[]{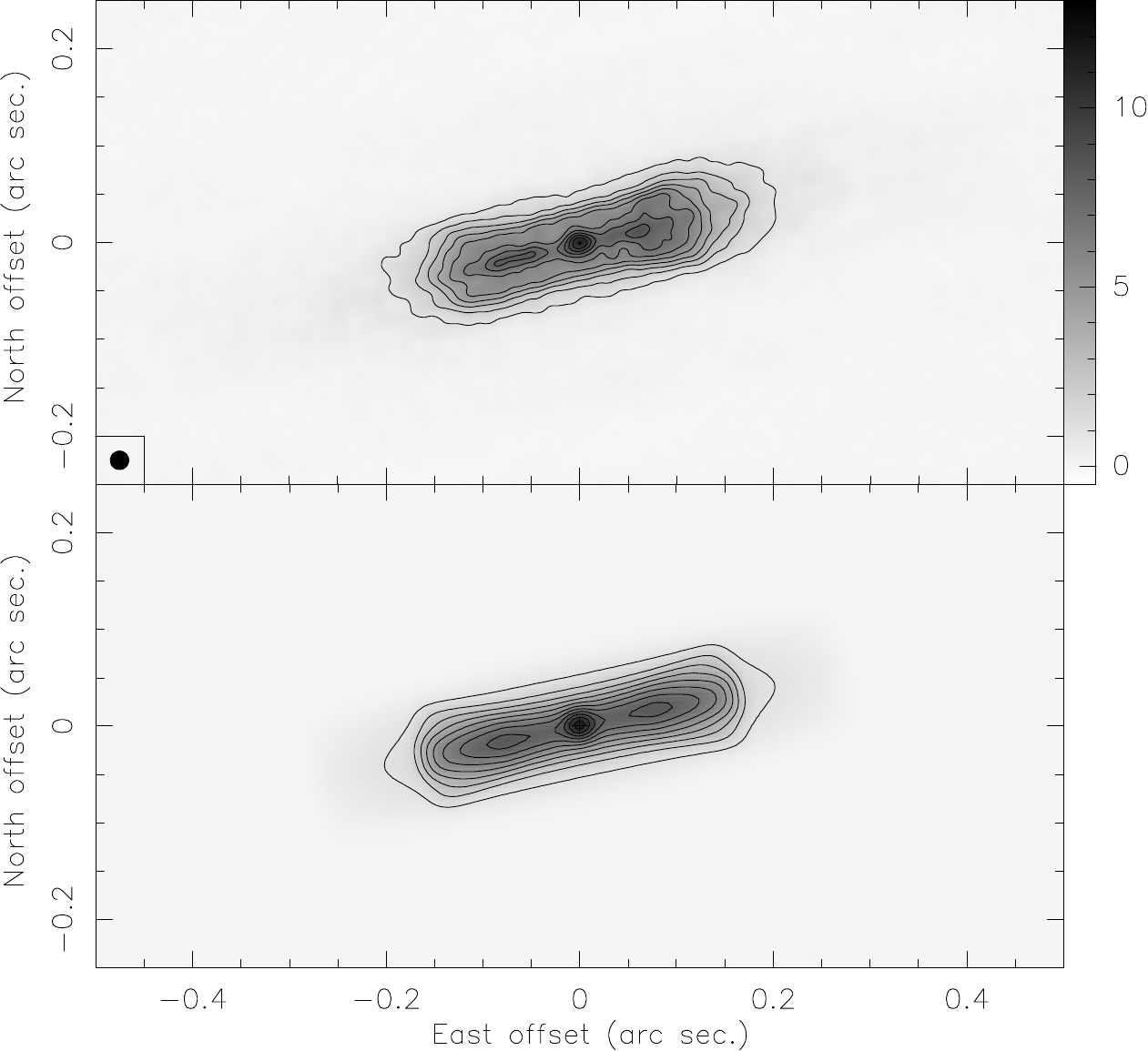}
}}
      \caption{Compact dust disk in the Red Rectangle. {\em Top:} ALMA 0.9 mm continuum map of the Red Rectangle. Contours are $-$1 and 1, 2, 3, ... \mJybeam; 1 \mJybeam\ is equivalent to 36.05 K in units of Rayleigh-Jeans equivalent brightness temperature. The clean-beam size at half maximum is (20$\times$20 mas) is shown in the inset. {\em Bottom:}
      Predicted continuum brightness distribution from our modeling (Sect.\ 3), all units and scales are the same as for the empirical map.
      }
         \label{1}
   \end{figure}

The result is similar if we compare the flux in our new data of \htrececn\ \jct\ line emission, $\sim$ 0.45 Jy with the old-data flux, $\sim$~0.4 Jy \citep{bujetal16}. The uncertainty may seem larger in the new maps due to the larger clumpiness triggered in the image reconstruction by the 30 times smaller beam, but we can also discard that a large amount of flux is filtered out in the new \htrececn\ maps. For our \trece\ \jtd\ data (to be presented in a future paper), the integrated flux profile is found to be very similar to that obtained in 2012, and only slightly smaller for the strongest peak at 0.5 \kms.
We have also compared the integrated fluxes of the old \doce\ \jtd\ data (with total peak flux of about 26 Jy), the new ones (total flux $\sim$ 25 Jy), and a single-dish profile obtained by \cite{debeck10} with the APEX telescope (for which we deduce a peak flux of about 30 Jy). The expected flux loss is \lsim\ 20\%\ for the very extended \doce\ emission.

In summary, the calibration uncertainty in these observations and in those published in 2013 is \lsim\ 10\%, and, despite the very high angular resolution, the amount of flux filtered out in the data presented in this letter is low, namely, $<$ 20\%.

From our continuum data, we calculate that the ratio between total continuum fluxes detected in both receiver subbands is $F$(344.5)/$F$(331.7) $\sim$ 1.089, which corresponds to a spectral index of $\alpha$ = 2.26$\pm$0.5 [with $\alpha$ being defined as $F$($\nu_2$)/$F$($\nu_1$) =  ($\nu_2$/$\nu_1$)$^\alpha$]. The uncertainty mainly comes from the errors in the spectral index derived for the phase calibrator,
by comparison with ALMA and NOEMA calibrator database as well as with the data obtained for the same calibrator in 2012. To improve our estimate, we collected data from cm- and mm-waves (see Fig.\ 2), including  measurements for dust relatively extended emission and the central free-free component. The fitting of the dust emission data yields a spectral index of 2.13, with a formal error of $\pm$0.04. The individual calibration errors are very small; in fact, they are smaller than the sizes of the symbols in this logarithmic graphic. The dust extended emission is in general calculated as the total flux minus the compact component. As a result, the derived fluxes can be somewhat dependent on subtraction procedures, in a way that is difficult to quantify. We underline the
good fitting of the data with a potential law, despite we are comparing very different observations, and the similar value obtained from fluxes in the ALMA subbands. We conclude that the spectral index of the dust emission ranges between 2.1 and 2.3.

      \begin{figure}[h]
   \centering\rotatebox{0}{\resizebox{7.5cm}{!}{
   \includegraphics[]{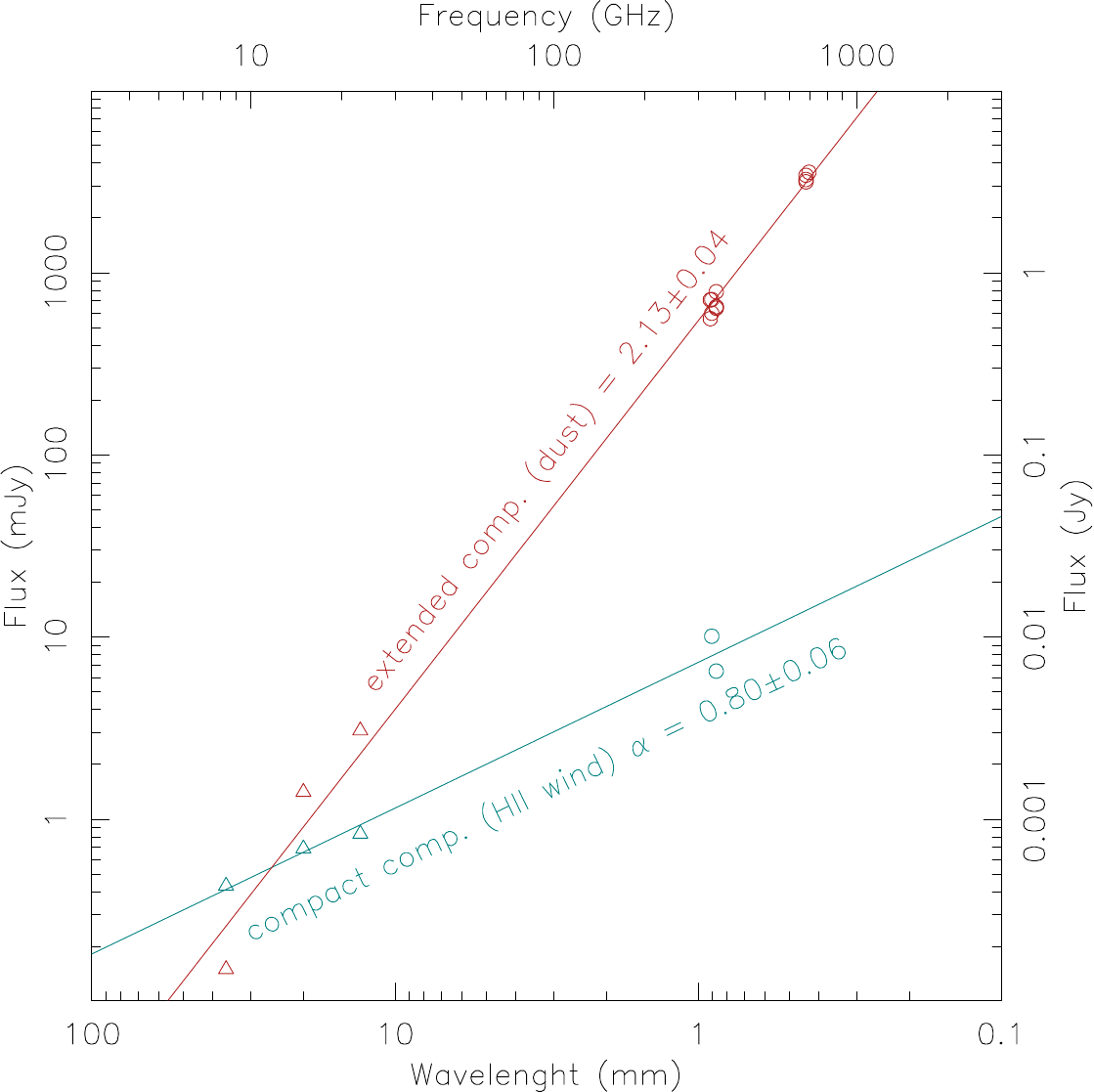}
}}
      \caption{Measurements of the continuum fluxes in the Red Rectangle, coming from the relatively extended dust emission (red) and the much more compact free-free emission (cyan). The cm-wave data (triangles) come from \cite{jura97}, submm measurements (circles) are from \cite{bujetal13b}, \cite{bujetal13b}, \cite{bujetal16}, and this paper. We also indicate the spectral indexes calculated from regression lines and the formal uncertainties.
      }
         \label{1}
   \end{figure}

   \section{Simple model of dust continuum emission}

\subsection{Dust grain properties}

In our brightness distribution modeling, the dependence of the dust intensity on the wavelength is  not really crucial, since only continuum maps at 0.9 mm are considered.
Because of the lack of detailed empirical information, we assume a very simple $\lambda$-dependence law of the absorption coefficient, $k$ \citep[see for instance][]{tielens05}:

$$
k(\lambda) = \pi r_g^2 n_g ~~{\rm if}~~ \lambda < \lambda_0 ~~~,
$$
$$
k(\lambda) = \pi r_g^2 n_g (\lambda/\lambda_0)^{-\beta}~~{\rm if}~~ \lambda > \lambda_0 ~~~.
$$
Here, $r_g$ is the grain typical radius ($\pi r_g^2$ is the geometric area offered to radiation) and $n_g$ is the number of grains per volume unit (cm$^{-3}$). $\beta$ represents a typical exponential dependence. We can assume that $\beta$ is equal to 1 (at least for not purely crystalline grains, for which $\beta$ can be higher), and $\lambda_0$ $\sim$ $10 \, r_g$.

If $\rho_g$ is the specific density of the solid material, the mass of this characteristic grain and the macroscopic density of dust material,
$\rho_d$(g\,cm$^{-3}$), are as follows: 
$$
m_g = \frac{4}{3}\pi r_g^3\rho_g  ~~~,~~~ \rho_d = n_g \frac{4}{3}\pi r_g^3\rho_g ~~~.
$$
 We will assume $\rho_g$ $\sim$ 3 g\,cm$^{-3}$ \citep[see e.g.,][]{csanchez98,tielens05}. The total dust mass in the disk is the integral of $\rho_d$ over the whole disk volume. We note that the conversion from absorption coefficient to dust mass density is approximately independent of the grain radius when grains are not very large ($r_g$ \gsim\ $\lambda/10$):
$$
\rho_d = \frac{4}{3} k(\lambda)\,\rho_g r_g ~~~~~~{\rm if}~~ \lambda < \lambda_0 = 10 r_g ~~~,
$$
$$
\rho_d = \frac{4}{3} k(\lambda)\, \rho_g r_g (\lambda/10\, r_g)^\beta ~\sim~ \frac{4}{3} k(\lambda)\, \rho_g r_g (\lambda/10\, r_g) ~~~~~~{\rm if}~~ \lambda > 10 r_g ~.
$$

We have estimated that dust emission at the relevant wavelengths follows a spectral index of $\alpha$ $\sim$ 2.1 -- 2.3 (see Sect.\ 2.2). In our simple dust emission description, such an index must be equal to 2 if the dust optical depth is very high, $\tau_d \gg 1,$ {\em } or the grains are very large, $r_g$ $\gg$ $\lambda$/10; while $\alpha$ = 3 in the limit of very low optical depth and small grains. Obviously, we are dealing with an intermediate case. Our value of $\alpha$ indicates $r_g$ \lsim\ 100 $\mu$m, the limit being attained in the case of optically thin dust emission. \cite{jura97} found, from mm- and cm-wave continuum observations, $r_g$ \gsim\ 200 $\mu$m. In view of the very simple descriptions used to deduce both limits, we consider that our result is compatible with that from \cite{jura97} and thus we adopted a characteristic grain radius of $r_g$ $\sim$ 150 $\mu$m.

The characteristic optical depth to get the departure of $\alpha$ with respect to the optically thick limit, must be $\tau_d$(0.9 mm) \lsim\ 2; otherwise, $\alpha$ \lsim\ 2.1 is expected even for very small grains (and $\alpha$ would be practically equal to 2 for relatively large grains, as deduced above).
We can get additional information on the opacity from the standard radiative-transfer formula of the brightness temperature in the Rayleigh-Jeans approximation (fully justified in our case), $T_B$ = $T_d (1-e^{-\tau_d})$, where $T_B$ is the observable brightness temperature and $T_d$ is a characteristic dust temperature; as mentioned (Sect.\ 1), this value cannot be larger than $\sim$ 1000 K. Our ALMA maps show relatively wide regions ($\pm$90 AU from the axis) with a brightness temperature of \gsim\ 200 K; therefore, we found a typical  value of $ \tau_d$ \gsim\ 0.2. From both limits, we deduce that $\tau_d$ $\sim$ 0.5-1 is a good approach in the central dense regions. This reasoning also leads to that the disk dust temperature must be high, that is, $\sim$ 500 K in those bright areas. Such dust temperatures may look too high compared with those in disks around young stars, but lower values are not compatible with the observed brightness level. In addition, we recall that high grain temperatures of this order have often been identified in the inner circumstellar dust around AGB and post-AGB stars \citep[e.g.,][]{men02, csanchez22} and that this region is probably a PDR (Sects.\ 1 and 4), where high temperatures are expected \citep[e.g.,][]{tielens05}.

We are aware that the above estimates are somewhat crude. But our subsequent modeling confirms (see Sect.\ 3.2) that the observations can be reproduced with values of the grain size, opacity, and temperature on the order of those deduced above.

\subsection{Radiative transfer treatment and data fitting}

Our simulations of dust emission are based on nebula models that are similar to those presented in previous papers on molecular lines from the Red Rectangle \citep[e.g.,][]{bujetal16}. Symmetry with respect to the axis and the equator is assumed. The main difference is that continuum emission is only detected from a much smaller region and does not show any trace of outflows leaving the disk. It is also relevant to note that the dust continuum offers no information on the kinematics and, therefore, the velocity field is not incorporated. We used the general properties of the dust absorption coefficients presented in Sect.\ 3.1. As in our previous works, our code produces brightness distributions by solving the radiative transfer equation for a large number of lines of sight. Emission from a small central HII region, simply defined by its total flux and extent, is also included in the brightness calculation to explain the central maximum. Finally, we obtain synthetic maps in physical units, directly comparable with observations, by convolving with the beam.

The free parameters are the disk shape and the distributions of dust density and temperature. The disk shape is assumed to be similar to the shape of the inner disk regions obtained by \cite{men02} and \cite{bujetal16}, with a thickness that increases with the distance to the symmetry axis. However, to explain the very narrow brightness distribution shown in our continuum maps, the dust disk must be significantly flatter than deduced in previous models. We also assumed the presence of a central region devoid of grains (where the HII region will be placed; see Sects.\ 1, 3.1) followed by a hot high-density component, beyond which the density rapidly falls down. We will see that this structure reproduces the observations, after fitting the scales and values of the physical conditions.

Our best-fit model is depicted in Fig.\ 3; the densities are given in cgs units (g cm$^{-3}$). We find that data fitting improves with a variation of thickness slightly faster than proportional and/or a small decrease of the density at distances \lsim\ 10$^{15}$ cm. The predicted map is shown in Fig.\ 1 (bottom), in which all scales and levels are the same as in the empirical data; we can see that the agreement with observations is satisfactory. As discussed in Sect.\ 3.1, there is little room for changes in the characteristic temperature and opacity values. In the determination of the dust material density given in Fig.\ 3, we assumed the typical values of the grain radius and solid material density discussed in Sect\ 3.1. From the density distributions in our nebula models, we derive a total mass in dust grains of 5$\times$10$^{-5}$ -- 10$^{-4}$ \ms, with an uncertainty due to model fittings compatible with the observations smaller than a factor of 1.5. Assumptions on the grain properties also yield uncertainties. The deduced total mass can in some cases vary proportionally to the assumed grain size (for large grains, Sect.\ 3.1). However, we have seen that the uncertainty in the characteristic grain radius must be moderate to explain the general observational features. In the case of very porous grains, the density of silicate or carbonaceous grain material can be significantly lower than our values: porous grains would imply smaller mass values. The best-fit dust temperature is of about 300-1000 K in the strongly emitting regions, from our previous discussion, slowly decreasing outward
(we assumed that $T_d$ is inversely proportional to the square root of the distance, according to radiative equilibrium) and with an upper limit of about 1000 K. Our best  model includes a small free-free central emission ($\sim$ 10 mas, $\sim$ 7 AU wide) with a flux of $\sim$7 mJy.

   \begin{figure}
   \rotatebox{0}{\resizebox{8.35cm}{!}{\includegraphics[]{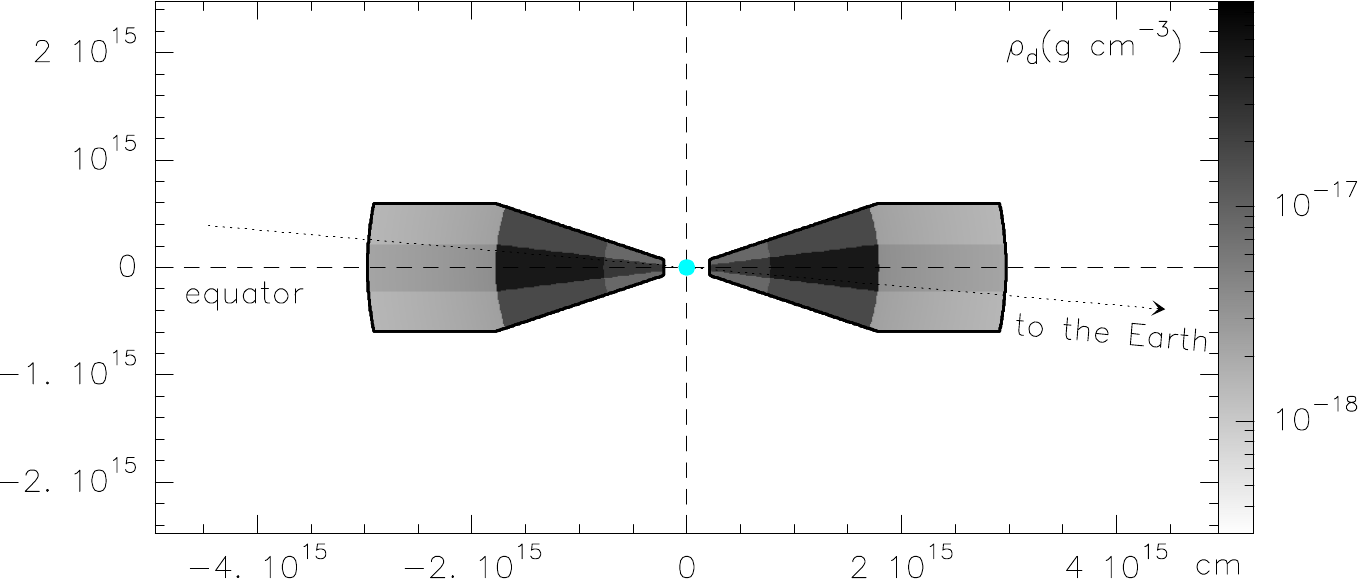}
}}
      \caption{Dust-rich disk model to fit our ALMA 0.9-mm continuum map in the Red Rectangle. The central cyan circle represents the 0.9-mm image of the inner HII region, whose shape is not well-known.}
         \label{1}
   \end{figure}
   
\section{Results and conclusions}

\subsection{Dust grain properties and total mass in dust}

We have confirmed previous indications \citep{jura95,jura97} of relatively large grains in the Red Rectangle. Our maps and modeling (Sect.\ 3) suggest typical dust grain radii as large as 150 $\mu$m, although the lack of information on the dust emission distribution at other wavelengths precludes a detailed discussion. We also suggests moderate, though non-negligible dust opacity in central disk regions.
The temperature of the dust-emitting region must be very high (Sect.\ 3.1), of several hundred K. This may look surprising, particularly in comparison with  disks around young stars, but lower temperatures would be incompatible with the very high brightness temperature detected in our maps (over $\sim$ 200 K in relatively wide regions) and are expected from previous studies of this source and similar ones (Sect.\ 3.1).

The dust density over our model nebula results in a total dust mass of  5$\times$10$^{-5}$ -- 10$^{-4}$ \ms\ (around 25 Earth masses). These values are compatible with the total gas mass deduced from molecular emission, 10$^{-2}$ -- 1.5$\times$10$^{-2}$ \ms\ \citep{bujetal16,gallardoc23}, for a standard dust-to-gas mass conversion of 100 -- 300 \citep[e.g.][]{tielens05,sahai99,csanchez98}.
We note that this gas mass value was derived by assuming a relative abundance $X$(\doce) \lsim\ 2$\times$10$^{-4}$, which is low compared with AGB stars and in accordance with the general lack of molecules found in these objects \citep{gallardoc22}.

\subsection{Dust and gas distribution: dust equatorial settlement}

Our continuum map (Fig.\ 1) shows that dust emission is strongly concentrated toward the equator and the center. We have seen that most of the dust emission, \gsim\ 90\% of the total flux (Sect.\ 2.2), comes from a very small central disk, with a diameter $\sim$ 0\farcs 45 ($\sim$ 5$\times$10$^{15}$ cm) and a width that is at least five times smaller. Our modeling (Sect.\ 3) confirms this result. We estimate, in view of the low brightness of outer regions and the slow decrease of the grain temperature (Sects.\ 2 and 3), that at least 80\%\ of the nebular dust is located in that flat disk.

The settlement of dust on central equatorial regions is a very robust result, particularly in comparison with the gas distribution. The \doce\ and \trece\ maps and the corresponding modeling, as per  \cite{bujetal13a, bujetal16} and our new maps (paper in preparation), show a significantly wider extent, with characteristic sizes around ten times larger than those of the dust.
Our models account for the continuum map by means of a strong decrease of grain density at about 2$\times$10$^{15}$ cm from the axis and 4$\times$10$^{14}$ cm (just 25 AU) from the equator (and a further decrease at distances that are about 50\% larger).
The molecular emission data and modeling indicate a decrease by a factor of 3.5 of the gas density at $\sim$ 10$^{16}$ cm \citep{bujetal16}, discarding the presence of significant amounts of material farther than 3 -- 4  10$^{16}$ cm. A similar concentration of dust in the equator, which has not been expected for gas and occurring simultaneously with grain growth, was recently found in another disk around a post-AGB binary \citep{corporaal23}. It also stands as a well-known property of disks in forming stars as a result of drag forces \citep[see, e.g.,][]{williams11,lin19}. We finally point out the somewhat asymmetric distribution of  dust emission from east to west (not reproduced by our symmetric model) and the intense central maximum that we attribute to a compact HII region (Sects.\ 3.2 and 4.4).

Our past and present molecular line maps do not show any evidence of strong changes in the molecular abundances close to the equator.
The molecular depletion towards the equator often found in disks around very young stars does not appear in the Red Rectangle, as expected, since such a phenomenon requires very low temperatures \citep[10 -- 20 K, see e.g.,][]{williams11} -- which are not present in our case.

   \begin{figure}
   \centering\rotatebox{0}{\resizebox{9cm}{!}{
   \includegraphics[]{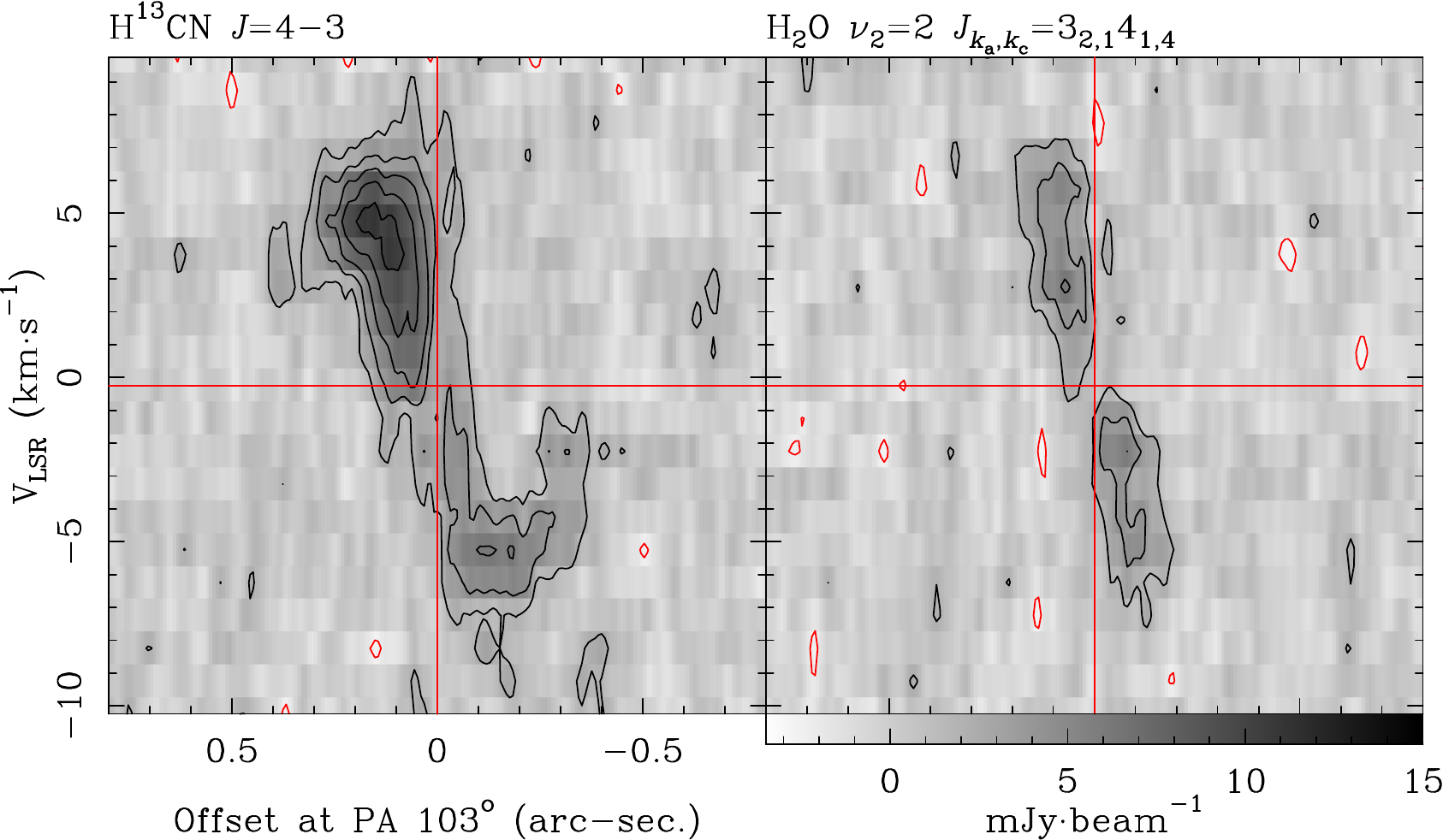}
}}
      \caption{Velocity vs.\ position plot along the equatorial direction (PA 103\degr) of the Red Rectangle nebula for 
      \htrececn\ \jct\ (left panel) and H$_2$O $\nu_2$=2 \jkk=\jkkul{3}{2}{1}{4}{1}{4} (right panel). The velocity resolution is 1\,\kms\ and the mean spatial resolution is 53 mas (see Sect. 2.1). Contours are $-$2 (red) and $+$2 to $+$10 by $+2$ (black) \mJybeam; 1 \mJybeam is equivalent to 3.6 K. The adopted LSR systemic velocity 
      ($-$0.25\,\kms) and center of the nebula (continuum peak) are indicated by the red lines.}
         \label{1}
   \end{figure}
   
\subsection{Emission of PDR-like lines: H$^{\it 13}$CN and H$_{\it 2}$O}

We have detected and mapped H$^{13}$CN \jct\ and H$_2$O $\nu_2$=2 \jkk=\jkkul{3}{2}{1}{4}{1}{4} (among other lines, see Sect.\ 2.1, to be discussed in a paper in preparation). Position-velocity diagrams of both lines are shown in Fig.\ 4; at least for H$^{13}$CN \jct, the Keplerian rotation signature is conspicuous.
\cite{bujetal16} argued that H$^{13}$CN is formed in an inner PDR in the Red Rectangle \citep[see also][]{thomas13}, that would occupy roughly about 120 AU and be also responsible for the CI and CII emission and the well-known presence of PAHs in the optical nebula. Our detection of H$_2$O confirms this result, as H$^{13}$CN and H$_2$O are thought to appear at the same time
only if stable molecules are partially photodissociated.
H$^{13}$CN \jct\ emission comes from a small region occupying $\sim$300 AU, larger than that deduced by \cite{bujetal16} from low-resolution mapping. The H$_2$O emission is more compact and closer to the previous PDR extent estimate. It is remarkable that the emission sharply disappears beyond a certain linear distance for both molecules,
indicating that they are only abundant within this inner region.

The extents of the dust disk and the gas rich in H$^{13}$CN and vibrationally excited H$_2$O are strikingly similar. It is clear that most dust continuum emission at 0.9 mm comes from that inner region, probably a PDR. This may explain the high dust temperatures required by our modeling, since temperatures in PDRs tend to be significantly higher than in surrounding regions (Sect.\ 3). The high energy required to excite H$_2$O line also indicates high excitation, perhaps including strong radiative pumping.

We stress that CO emission, which shows a very large total extent, is also prominent in inner disk regions within about $\pm$0\farcss 5 from the center: stable molecules, at least CO and H$_2$O, are abundant in the gas responsible for the CI, CII, and \htrececn\ emissions or in a tightly associated region.
The PDR and dust-emitting region are practically coincident and placed between the very compact HII region and the outer gas emitting in CO lines (where photodissociation would be negligible). We recall that the profiles of CI and CII (obtained with low angular resolution) show, together with wings coming from the inner disk rotating at high velocity, a peak at the systemic velocity whose origin remains unknown.

\subsection{Central HII region and free-free emission}

We detect a central continuum source, which is very compact and practically unresolved. Its extent, 10 -- 15 mas (7 -- 11 AU), is smaller than the inner dust boundary ($\sim$ 40 mas in diameter; see Sects.\ 1, 3), but larger than the binary orbit size ($\sim$ 3 mas, 2 AU). It very probably corresponds to the counterpart at shorter wavelengths of the ionized component already  mapped by  \cite{jura97}. We measured its total flux at $\lambda$ = 0.9 mm, 7 -- 8 mJy, both from subtraction from the overall continuum maps and model fitting. Given the uncertainties of the very different observations (Sect.\ 2.2) and our poor knowledge on the HII-region properties, this flux value is compatible with the extrapolation of the previous fluxes following the spectral slopes expected for emission from an ionized wind \citep[e.g.][]{panagia75}. The smaller extent of our free-free emission image is also expected.

\begin{acknowledgements}
This work is part of the I+D+i project EVENTs/Nebulae Web, PID2019-105203GB-C21 and PID2019-105203GB-C22,
funded by the Spanish MCIN/AEI/10.13039/501100011033. ALMA is a partnership
  of ESO (representing its member states), NSF (USA) and NINS (Japan),
  together with NRC (Canada) and NSC and ASIAA (Taiwan), in cooperation
  with the Republic of Chile. The Joint ALMA Observatory is operated by
  ESO, AUI/NRAO and NAOJ.
\end{acknowledgements}


\end{document}